\documentclass[aps,reprint,twocolumn]{revtex4-2}  

\usepackage{graphicx}
\usepackage{amsmath}
\usepackage{graphicx}
\usepackage{dcolumn}
\usepackage{bm}
\usepackage{float}
\usepackage{xcolor}
\usepackage{soul} 
\usepackage{siunitx}
\usepackage{amsmath}
\usepackage{hyperref}

\usepackage{hyperref}
\begin{document}



\title{Learning Thermal Response Forces: A Method for Extending the Thermodynamic Transferability of Coarse-Grained Models via Machine-Learning}

\author{Patrick G. Sahrmann}
\email{sahrmann@lanl.gov}
\author{Benjamin T. Nebgen}
\author{Kipton Barros}
\author{Brenden W. Hamilton}

\affiliation{Theoretical Division, Los Alamos National Laboratory, Los Alamos, New Mexico 87545, USA}

\date{\today}

\begin{abstract}

Machine-learned (ML) coarse-grained (CG) models are a promising tool for significantly enhancing the efficiency of molecular simulations by systematically removing degrees of freedom while retaining fidelity to the underlying fine-grained model.
The CG potential of mean force (PMF) is inherently dependent on thermodynamic conditions and, hence, a CG force-field (FF) which is trained at one thermodynamic state point is not necessarily accurate at another.
We propose, in this work, a novel and data-efficient means of learning temperature dependence into ML CG force-fields via training on the thermal response forces of the PMF.
We demonstrate how incorporating these terms into ML CG FFs confers significantly improved transferability for CG water models and demonstrate how this transferability enables accurate and predictive CG dynamics. 

\end{abstract}

\maketitle

A grand challenge in computational physics is the development of general procedures by which physics at one spatiotemporal scale can be mapped to a coarser scale, often referred to as scale bridging.
Molecular coarse-grained (CG) modeling is considered one of the first rungs on the scale-bridging `ladder' in which atomistic level models are mapped to a coarser resolution,
such that the CG degrees of freedom may comprise the behavior of upwards of hundreds of atoms, accelerating simulation efficiency by orders of magnitude.\cite{teza2020exact,doi:10.1021/acs.jctc.2c00643,curtarolo2002dynamics,swinburne2023coarse}

The Mori-Zwanzig (MZ) formalism provides an exact mapping of the time evolution of the CG degrees of freedom as stemming from the CG potential of mean force (PMF), a conservative FF which retains all static equilibrium behavior of a CG system, 
and a non-conservative memory and noise term which ensures accurate CG dynamics.\cite{10.1063/1.1731409} 
In recent years, machine-learned (ML) methods have shown significant promise in learning CG FFs which can recapitulate both equilibrium and dynamical behavior, as described by the PMF.\cite{PhysRevLett.131.177301,SAHRMANN2025102972,doi:10.1021/acs.jctc.2c00616,doi:10.1021/acs.jctc.4c00788,DURUMERIC2023102533,doi:10.1021/acs.jpcb.3c05928,Duschatko2024,Charron2025,majewski2023machine,10.1063/5.0274785,doi:10.1021/acs.jctc.2c00706,10.1063/5.0120386}

The primary challenge in CG model development, which precludes their widespread adoption, is model transferability. 
Unlike atomistic FFs, the CG PMF is not a potential but a free energy, and therefore varies at each thermodynamic state point.
Consequently, CG models must incorporate state-point dependence for reasonable transferability.\cite{Lu2011TheMC,parsa2020large}  

Development of CG FFs which are explicitly temperature dependent, in particular ML CG FFs, remains a largely unexplored area. 
A significant challenge in learning temperature transferable CG FFs is the absence of theoretical methods which enable explicit training of temperature dependence.\cite{duschatko2024thermodynamicallyinformedmultimodallearning}
Current methods rely on learning a PMF interpolation across multiple thermodynamic state points, attempting to implicitly learn the temperature dependence.
However this approach remains costly, while offering no explicit quantitative estimate of the accuracy of the temperature dependence of the PMF.\cite{10.1063/5.0022431}

Here, we propose the first theoretical methodology wherein the temperature dependence of the CG PMF can be probed explicitly from the forces of the PMF.
Additionally, this constitutes the first work, to our knowledge, where thermodynamic quantities such as the entropy missing due to atomistic information loss \cite{doi:10.1021/acs.jpclett.9b01228} can be learned explicitly from single state point data. 
We subsequently demonstrate how ML models can be trained from these thermal response forces to learn a CG PMF from a single thermodynamic state point which varies smoothly, accurately, and in a physically-interpretable manner with temperature.

To begin, we assume a set of fine-grained (FG) degrees of freedom, $\mathbf{r}^n$, which interact according to a potential, $u(\mathbf{r}^n)$.
We also assume that this system is under thermodynamic equilibrium at a temperature, $T$. The CG degrees of freedom are described as a subset of the original degrees of freedom, $\mathbf{R}^N$, such that $N < n$,
where the mapping operator, $\mathcal{M}$, bridges the FG and CG descriptions.
The PMF, which describes the static equilibrium behavior of this CG set, $F(\mathbf{R}^N;T)$, is defined up to a constant as\cite{10.1063/1.2938860}
\begin{equation}\label{Eq. 1}
    F(\mathbf{R}^N;T) \propto -k_BT \cdot \ln \int \text{d}\mathbf{r}^n e^{-u(\mathbf{r}^n)/k_BT} \delta(\mathcal{M}(\mathbf{r}^n) - \mathbf{R}^N).
\end{equation} 

Currently, the most common means of training ML CG FFs is via force-matching.\cite{doi:10.1021/jp044629q,10.1063/1.2938857,doi:10.1021/acscentsci.8b00913} 
Consider the force on CG site $I$, $\mathcal{F}_I(\mathbf{R}^N;T) \equiv -\nabla_I F(\mathbf{R}^N;T)$. 
It can be shown that the CG forces are related to the conditional expectation of the FG forces, $f_i(\mathbf{r}^n)$, such that 
\begin{equation}\label{Eq. 2}
    \mathcal{F}_I (\mathbf{R}^N;T)  = \langle \Xi^i_I f_i(\mathbf{r}^n)\rangle_{\mathbf{R}^N, T},
\end{equation}
where $\langle\cdot\rangle_{\mathbf{R}^N}$ signifies the expectation of a variable conditioned at a point in CG space, $\mathbf{R}^N$, and Einstein summation notation is implied over atom indices. 
The choice of force-mapping operator, $\Xi$, is constrained by the choice of mapping operator.\cite{doi:10.1021/acs.jpclett.3c00444} 
By regressing a CG FF on the forces of Eq.~\eqref{Eq. 2}, one ideally obtains the CG PMF at a given temperature. 
However, the temperature dependence of the PMF, as shown in Eq.~\eqref{Eq. 1}, is highly non-trivial, which hinders the transferability of this learned PMF across different temperatures. 
Temperature weights how FG microstates contribute to a given CG macrostate, and consequently, a CG model trained to reproduce a single PMF will not retain accurate behavior beyond temperatures where this state weighting differs significantly.

The assumption here, which has been employed previously\cite{Lu2011TheMC}, is that the CG free energy varies slowly enough with temperature such that entropy is the dominant contribution to the temperature dependence of the free energy. 
To this end, we consider the PMF at a single reference temperature, $T_0$ and Taylor expand to define $F(\mathbf{R}^N;T)$. Then, to second order,
\begin{equation}\label{Eq. 3}
\begin{split}
    F(\mathbf{R}^N;T) &= F(\mathbf{R}^N;T_0) - \Delta T \cdot S(\mathbf{R}^N;T_0)\\
    & - \frac{\Delta T^2}{2T_0} \cdot C_V(\mathbf{R}^N;T_0) + \mathcal{O}(\Delta T^3),
\end{split}
\end{equation}
where $\Delta T = T- T_0$, and it is implied that all other extensive variables are held constant in the expansion. 
$S(\mathbf{R}^N;T_0)$ and $C_V(\mathbf{R}^N;T_0)$ constitute the CG entropy and heat capacity, respectively. 
We provide justification in Section I of the Supplemental Material (SM) (see \cite{supplemental} with references \cite{THOMPSON2022108171, doi:10.1021/acs.jpcb.3c04473, https://doi.org/10.1002/jcc.21787,https://doi.org/10.1002/wics.1460} therein) for when termination of this Taylor series is acceptable. 
Assuming the temperature expansion of Eq.~\eqref{Eq. 3} is valid, one must obtain a representation of the entropic component of the CG PMF to enable accurate transferability of the CG model beyond the temperature at which force data was acquired. 
The entropy, however, is practically incalculable for non-trivial systems.

Here, we circumvent this typically formidable complication by recognizing that, while the explicit entropy is not a readily calculable quantity, the entropic component to the CG PMF can be trained in exactly the same manner as the PMF is trained via its forces, which are calculable. 
Let $\mathcal{S}_I$ denote the entropic force on CG site $I$, i.e. $\mathcal{S}_I(\mathbf{R}^N;T) \equiv -\nabla_IS(\mathbf{R}^N;T)$. It can be shown that 
\begin{equation}\label{Eq. 4}
\begin{split}
    \mathcal{S}_I(\mathbf{R}^N;T) &= -\frac{1}{k_BT^2} \cdot \Bigl ( \langle u(\mathbf{r}^n) \cdot \Xi^i_I f_i(\mathbf{r}^n) \rangle_{\mathbf{R}^N, T}\Bigr.\\
    &\Bigl. - \langle u(\mathbf{r}^n)\rangle_{\mathbf{R}^N, T} \cdot \langle \Xi^i_I f_i(\mathbf{r}^n) \rangle_{\mathbf{R}^N, T} \Bigr ) ,
\end{split}
\end{equation}
where the entropic force constitutes a mixed second-order derivative of the CG free energy, allowing it to be represented as a covariance. 
We note that Eq.~\eqref{Eq. 4} provides predictive capability on the temperature dependence near a thermodynamic state point. 
I.e., entropic effects are directly related to the correlation between the energy and mapped forces,
which may become large for severely coarse mappings or when long-range order is present. 

The heat capacity, unlike the entropy, is calculable from restrained MD simulations
\begin{equation}\label{Eq. 5}
    C_V(\mathbf{R}^N;T) = \frac{1}{k_BT^2} \cdot\left ( \langle u^2(\mathbf{r}^n)\rangle_{\mathbf{R}^N, T} - \langle u(\mathbf{r}^n)\rangle^2_{\mathbf{R}^N, T} \right ).
\end{equation}
Furthermore, the heat capacity `force' on CG site $I$, $\mathcal{C}_I(\mathbf{R}^N;T) \equiv -\nabla_I C_V(\mathbf{R}^N;T)$, is also calculable (see SM Section I).


We note that, while it is formally equivalent to regress the CG FF onto the mapped atomistic forces in the atomistic ensemble to learn the CG PMF, no such reinterpretation is possible for the entropic and heat-capacity forces.
This is directly attributable to the differing statistical interpretations of the PMF force and entropic/heat capacity forces, where the former is a conditioned average and the latter are conditioned higher-order force-energy moments. 
We emphasize that the theoretical results presented not only enable immediate construction of a transferable CG FF via the Taylor series of Eq.~\eqref{Eq. 3}, but also enable one to obtain $\dfrac{\partial F}{\partial \mathbf{R}}, \dfrac{\partial^2 F}{\partial \mathbf{R} \partial T}$ and $\dfrac{\partial^3 F}{\partial \mathbf{R}\partial T^2}$, allowing for any functional form of the temperature-dependence to be explicitly learned. 

\begin{figure}[htpb]
  \includegraphics[width=0.4\textwidth]{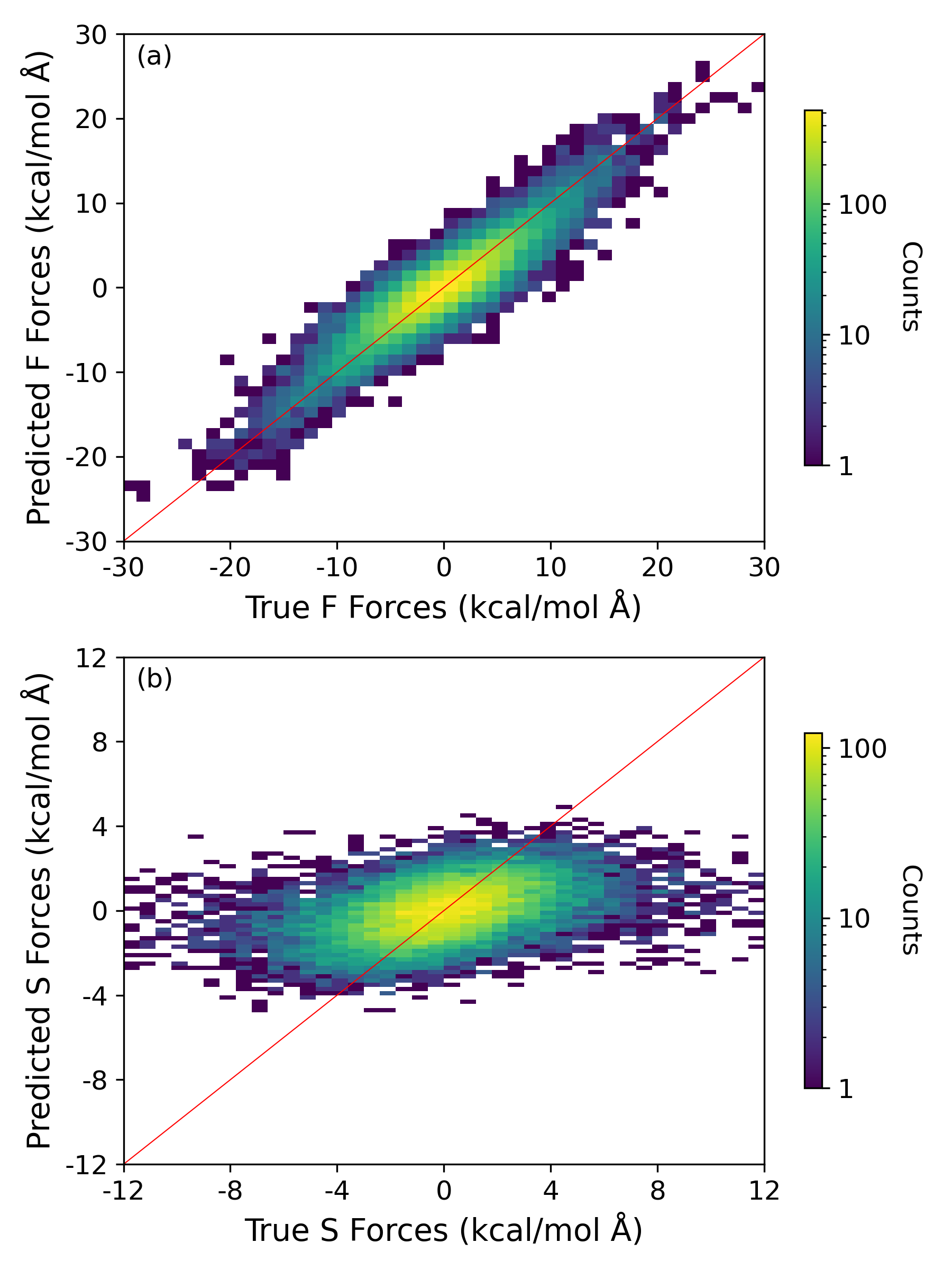}
  \caption{Parity plots for training of (a) PMF forces and (b) entropy forces. Forces are scaled by $T_0$ for dimensional equality between all quantities shown.}
  \label{fig:parity}
\end{figure}

To demonstrate the utility of the approach presented, we consider simulations of water across varying temperatures. 
All water simulations consisted of 528 water molecules at a density of $\rho = 1 \ \text{gm}/\text{cm}^3$, parameterized via the SPC/Fw water model.\cite{10.1063/1.2136877} 
The all-atom (AA) to CG mapping consisted of mapping each water molecule to its center of mass. The force-mapping operator was unity for all atoms involved in a CG site. 

We employ the Hierarchically Interacting Particle Neural Network with Tensor Sensitivity information (HIP-NN-TS) for ML modeling of the CG PMF, CG entropy, and CG heat capacity.\cite{10.1063/1.5011181, 10.1063/5.0142127} 
Information regarding training of HIP-NN-TS models and CG simulations are provided in Section II of the SM. 
CG models were trained at 250, 300, and 350 K. $0^{\text{th}}$, $1^{\text{st}}$, and $2^{\text{nd}}$ order terms were included according to Eq.~\eqref{Eq. 3}. 
Results for the 250 K and 350 K CG models are shown in Section II of the SM.
Force parity plots are shown in Fig.~\ref{fig:parity} for all order terms, and force metric data is included in Section II of the SM. 
We note that the $1^{\text{st}}$ order modeling shares theoretical similarities to an energy-entropy decomposition approach towards learning transferable CG models.\cite{Kidder2021,10.1063/1.5094330,10.1063/5.0281108} 
We provide an additional comparison to this approach in Section III of the SM.

From Fig.~\ref{fig:parity}, a hierarchy in force magnitudes is observed and are well fit by the model.
At the reference temperature and comparable $\Delta T$, PMF forces are greater in magnitude than entropic forces, which are greater than heat capacity forces. 
Hence, the forces of a CG model are largely controlled by the PMF at small $\Delta T$ and by the entropy and heat capacity at large $\Delta T$.  
It is clear from Fig.~\ref{fig:parity} that as higher-order temperature dependence is probed, the signal-to-noise of the resulting force contributions diminishes. 
Consequently, while all terms in the Taylor series of Eq.~\eqref{Eq. 3} can be calculated from restrained MD at any order (see Section I of SM), the degree of each moment, and hence the moment's variance, 
increases such that probing higher-order terms at a given temperature is unlikely to be beneficial.

Fidelity of all CG models was assessed according to recapitulation of structural correlations. We refer to CG models which contain $0^{\text{th}}$, $1^{\text{st}}$, and $2^{\text{nd}}$ temperature-dependence as CG, TCG1, and TCG2 models, respectively.
\begin{figure}[htpb]
  \includegraphics[width=0.4\textwidth]{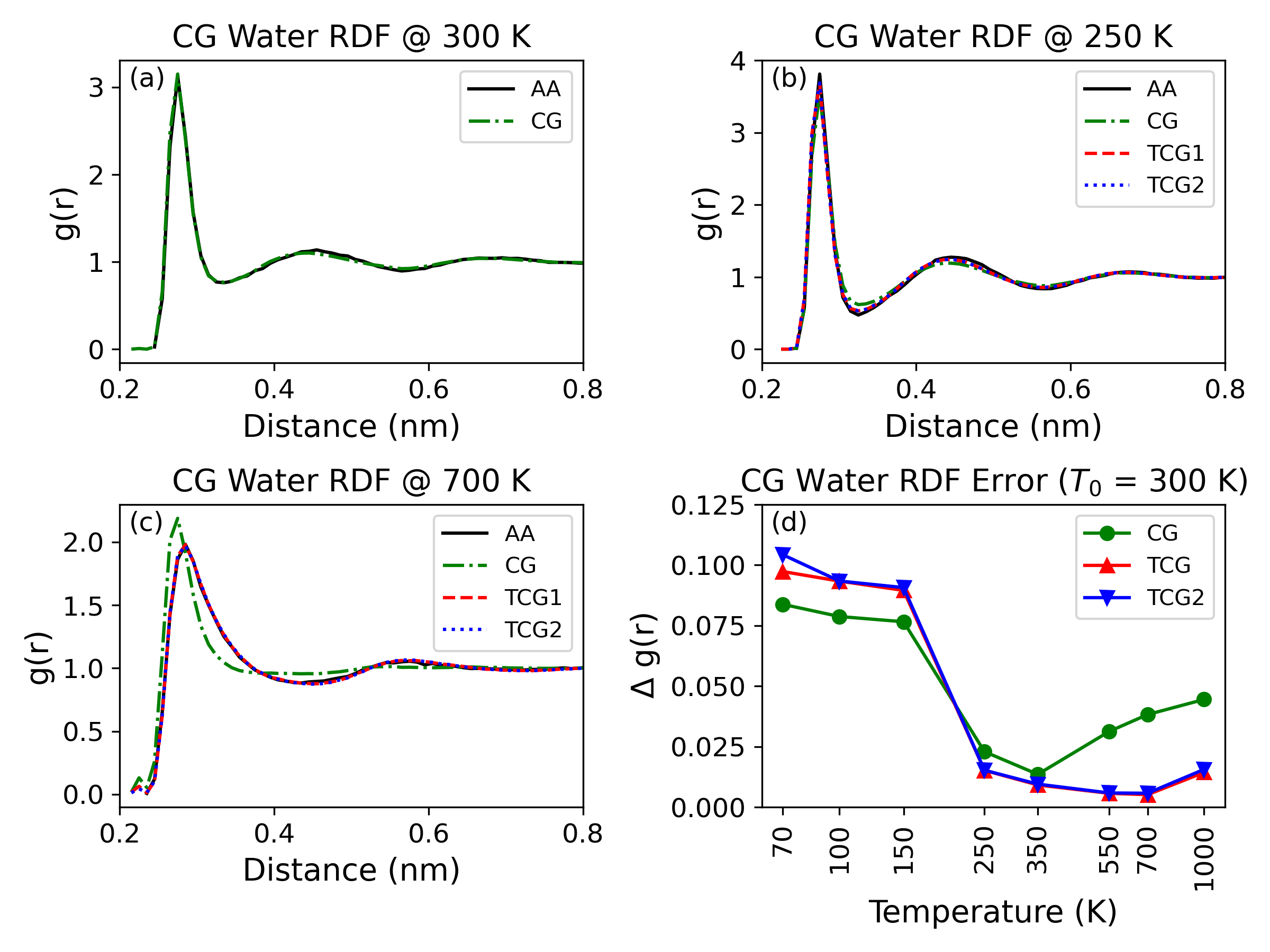}
  \caption{RDF of CG models ($T_0 = 300$ K). RDFs of mapped AA and CG models at (a) trained temperature, (b) T = 250 K, and (c) T = 700 K are shown. (d) RDF error across simulated temperatures for all CG models.}
  \label{fig:rdf}
\end{figure}
Representative radial distribution functions (RDFs) and RDF errors are shown in Fig.~\ref{fig:rdf}. 
RDF error was quantified by $\Delta g(r) = \frac{1}{r_{max}}\int^{r_{max}} \text{d}r |g(r) - g_\text{AA}(r)|$ where ${r_{max}}$ = 1.2 nm. 
It is clear that there exists a $\Delta T$ regime spanning hundreds of Kelvin wherein the addition of temperature correction terms greatly improves the accuracy of the CG model. 
However, after some threshold of $\Delta T$, addition of temperature correction terms is not beneficial for model fidelity. 

Notably, there is an asymmetry in this threshold $\Delta T$, i.e., the benefits of temperature corrections to the CG PMF via Eq.~\eqref{Eq. 3} diminishes at a faster rate for $\Delta T < 0$ relative to $\Delta T > 0$. 
This inaccuracy is likely a consequence of the Taylor series approximation of Eq.~\eqref{Eq. 3}.
Conversely, the threshold of improved accuracy for  $\Delta T > 0$ is hundreds of Kelvin, which is commensurate with the assumption that the CG PMF varies slowly. 
Representative angular distribution functions (ADFs) of triplets within the  $1^{\text{st}}$ solvation shell and ADF errors, quantified by $\Delta P(\theta) =  \frac{1}{\theta_{max}} \int^{\theta_{max}} \text{d}\theta |P(\theta) - P_\text{AA}(\theta)|$ where ${\theta_{max}} = 180^\circ$, 
are shown in Fig.~\ref{fig:adf} of the SM, from which largely the same conclusions regarding transferability can be made.

\begin{figure}[htpb]
  \includegraphics[width=0.4\textwidth]{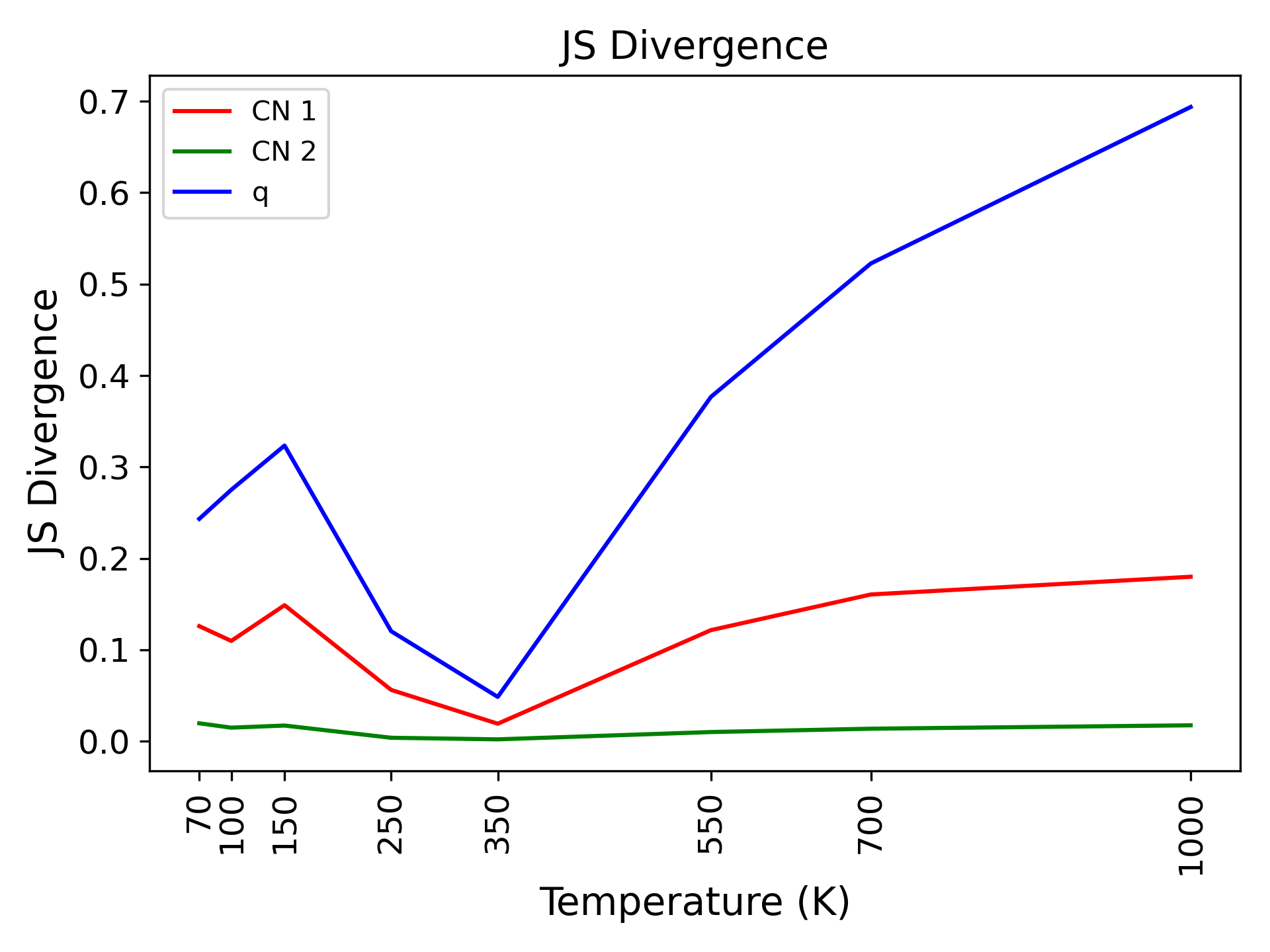}
    \caption{Change in local environment for mapped AA model relative to 300 K, assessed via the JS divergence between order parameter distributions at a given temperature and 300 K. Order parameters consist of coordination numbers within first and second solvation shell and tetrahedral order parameter.}
  \label{fig:env}
\end{figure}

We next examine how the local structural environments change with temperature for both AA and CG models. 
We employ coordination numbers (CNs) and the tetrahedral order parameter, $q = 1 - \frac{3}{8} \sum_{j=1}^3\sum_{k=i+1}^4(\cos (\psi_{jk}) + \frac{1}{3})^2$ as representative order parameters which describe the local CG environment. 
We quantify how dissimilar the local CG water environments are at other temperatures, with respect to the temperature employed for training ($T_0 = 300$ K), via the Jensen-Shannon (JS) divergence for the distributions of these afore-mentioned order parameters, as shown in Fig.~\ref{fig:env}. 
We provide further information on these order parameter distributions for the CG models developed in Fig.~\ref{fig:ops} of the SM. 
Intriguingly, we find that the model accuracy of the CG models developed is not strictly correlated with how dissimilar the local environment is with respect to the training temperature. 
E.g., the model error at 70 K is much larger than at 550 K despite their structural dissimilarities being roughly equal. This implies that the transferable CG models are capable of genuine extrapolation, as seen for $\Delta T > 0$, 
and further suggests that the breakdown in accuracy for $\Delta T < 0$ is a consequence of the assumption that the CG PMF various slowly with $T$ in this regime.

Lastly, we investigate the dynamical behavior of the CG models presented. 
Diffusion plots are shown in Fig.~\ref{fig:diff}(a).
We find that in general, the dynamics of the CG models are faster than the mapped AA model, commensurate with previous findings,\cite{10.1063/5.0212973} and can be directly attributed to the lack of MZ frictional terms during CG simulation. 
However, the addition of temperature correction terms to the CG PMF lead to temperature dependence in the activation energy (deviations from Arrhenius behavior), recovering the non-linear trends of the AA diffusion curve
in the TCG curves, an aspect wholly missing from the CG model.

We note that, under the assumption of overdamped Kramers' theory,\cite{KRAMERS1940284} diffusion exhibits the following relationship,
\begin{equation}\label{Eq. 6}
    D(T) = A(T) e^{-F^\dagger / k_BT}.
\end{equation}
Eq.~\eqref{Eq. 6} cleanly separates diffusion into a thermodynamic factor $e^{-F^\dagger / k_BT}$, 
and a largely dynamical factor $A(T)$ (see Section IV of the SM). 
Importantly, simulations of the CG PMF ideally will capture the free energy barrier, but will not capture $A(T)$ unless the missing AA frictional forces are accounted for.\cite{10.1063/5.0212973}
Our results are indicative of this trend, i.e., incorporation of $T$-dependent terms correctly introduces non-Arrhenius behavior but does not bridge the gap in magnitudes between CG and AA diffusion.

\begin{figure}[htpb]
  \includegraphics[width=0.4\textwidth]{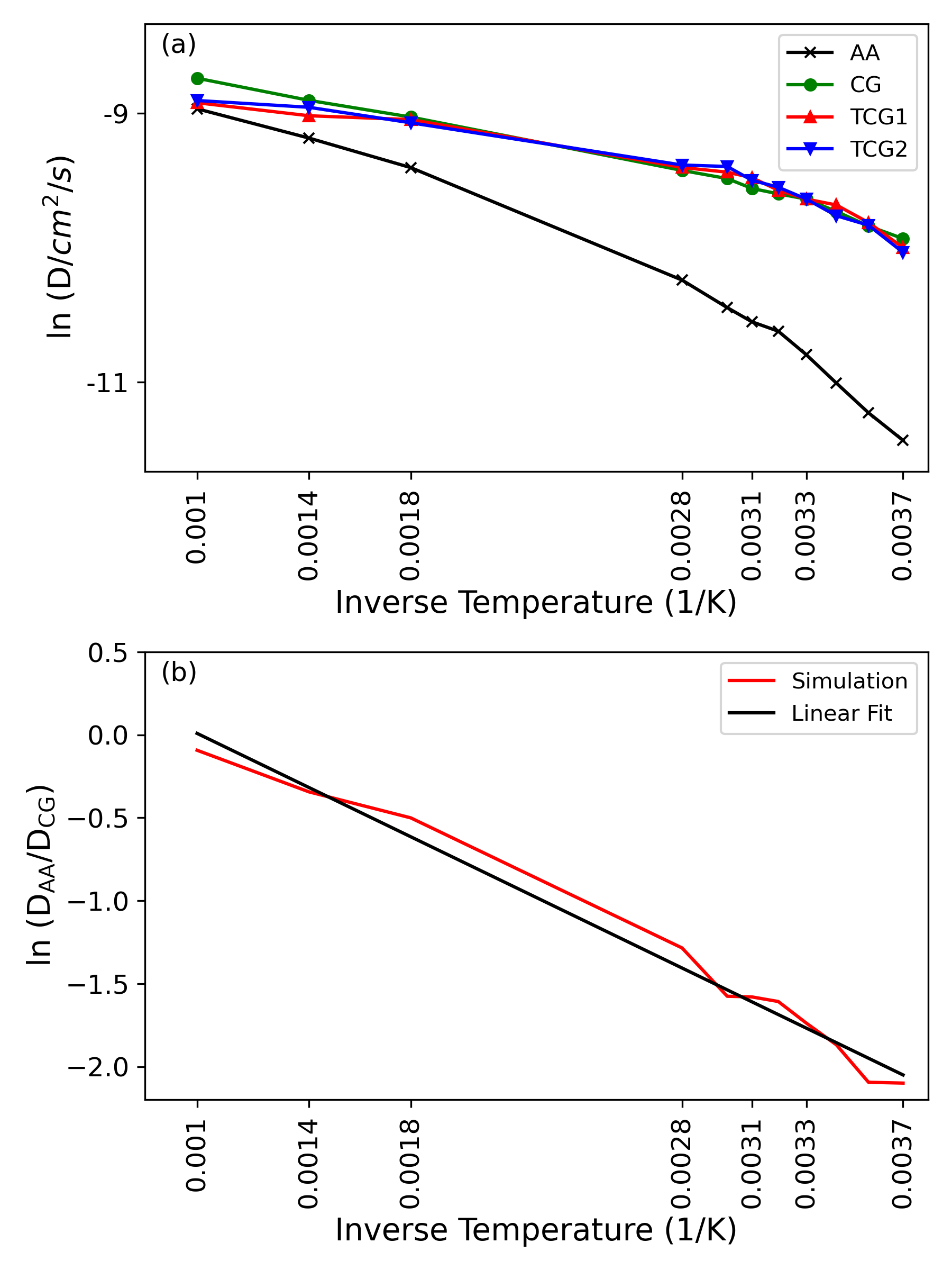}
  \caption{(a) Diffusion behavior across temperature for all models are shown. (b) Linear fit of the log ratio of diffusion constants between mapped AA and TCG2 model as a function of $T$ is shown.}
  \label{fig:diff}
\end{figure}

However, assuming both AA and TCG models ideally share the same free energy across temperatures, any disparity in diffusion can be  attributed to $A(T)$. 
Remarkably, we find that this difference in $\log A(T)$ follows a linear relationship, as shown in Fig.~\ref{fig:diff}(b). 
As we show in Section IV of the SM, this allows one to infer both correct CG dynamics from CG PMF simulations and a quantitative relationship for how time is re-normalized with $T$.
These results constitute a new paradigm in CG modeling in which transferable CG models predict CG dynamics by enabling a direct method
to correct rate dependent phenomena in post, grounded directly in the separation of MZ and non-MZ terms.

In conclusion, we have developed a means of learning the temperature-dependence of the CG PMF by regression of ML CG FFs on its thermal response forces. 
Learning temperature-dependence in this manner confers significantly enhanced temperature transferability while training from only a single thermodynamic state point. 
We expect that incorporation of multiple thermodynamic state points within the workflow presented will further improve thermodynamic transferability of ML CG FFs by enabling learning of thermal response forces at different 
temperatures. 
The presented work also can be readily applied to training of CG FFs where temperature dependence is not assumed to follow a Taylor series expansion but instead learned from training, in which the procedure elaborated in this work constitutes a Sobolev training scheme.\cite{NIPS2017_758a0661} 
In particular, training across temperature regimes where large structural reorganization is observed, such as at a phase transition, will be pivotal in smoothly interpolating ML CG FFs across phases.

\begin{acknowledgments}
Funding for this project was provided by the Advanced Simulation Computing (ASC) Program through the Nicholas C. Metropolis Postdoc Fellowship in Computational and Computer Sciences. Partial funding was provided by the Advanced Simulation and Computing Physics and Engineering Models project (ASC-PEM). This research used resources provided by the Los Alamos National Laboratory (LANL) Institutional Computing Program. This work was supported by the U.S. Department of Energy (DOE) through LANL, which is operated by Triad National Security, LLC, for the National Nuclear Security Administration of the U.S. Department of Energy (Contract No. 89233218CNA000001). Approved for unlimited release: LA-UR-26-21389.
\end{acknowledgments}

\bibliographystyle{apsrev4-2}
\bibliography{abbrev,references}


\clearpage
\onecolumngrid  

\setcounter{table}{0}
\setcounter{figure}{0}
\renewcommand{\thetable}{S\arabic{table}}
\renewcommand{\thefigure}{S\arabic{figure}}

\setcounter{footnote}{0}
\renewcommand{\thefootnote}{\fnsymbol{footnote}}

\begingroup
\centering
{\large\bfseries Supplementary Material\par}
\vskip 0.5em
{\large\bfseries for “Learning Thermal Response Forces: A Method for Extending the Thermodynamic Transferability of Coarse-Grained Models via Machine-Learning”\par}
\vskip 0.5em
{\normalsize Patrick G. Sahrmann, Benjamin T. Nebgen, Kipton Barros, and Brenden W. Hamilton\par}
{\itshape Theoretical Division, Los Alamos National Laboratory, Los Alamos, New Mexico 87545, USA \par}
\endgroup




\author{Patrick G. Sahrmann}
\email{sahrmann@lanl.gov}
\author{Benjamin T. Nebgen}
\author{Kipton Barros}
\author{Brenden W. Hamilton}

\affiliation{Theoretical Division, Los Alamos National Laboratory, Los Alamos, New Mexico 87545, USA}

\date{\today}

\title{Supplemental Material for Learning Thermal Response Forces: A Method for Extending the Thermodynamic Transferability of Coarse-Grained Models via Machine-Learning}

\maketitle

\section{Coarse-Grained Thermodynamic Identities}
We elaborate first on useful statistical mechanical identities and definitions. We note that for any function of the AA configurational space, $A(\mathbf{r}^n)$, the following is true
\begin{equation}\label{Eq. S1}
\dfrac{\partial \langle A(\mathbf{r}^n)\rangle_{\mathbf{R}^N}}{\partial T} = \frac{1}{k_BT^2} \cdot \Big (\langle A(\mathbf{r}^n)\cdot u(\mathbf{r}^n)\rangle_{\mathbf{R}^N}  - \langle A(\mathbf{r}^n)\rangle_{\mathbf{R}^N} \cdot \langle u(\mathbf{r}^n)\rangle_{\mathbf{R}^N} \Big )\tag{S1}
\end{equation}
Clearly, $A(\mathbf{r}^n) = \Xi_I f(\mathbf{r}^n)$ provides the direct evaluation of the entropic force. Next, we denote $\delta A(\mathbf{r}^n) = A(\mathbf{r}^n) - \langle A(\mathbf{r}^n) \rangle_{\mathbf{R}^N}$. We define the following centered moment function, $M(\alpha)$, as
\begin{equation}\label{Eq. S2}
M_I(\alpha) = \langle \delta \Xi^i_I f_i(\mathbf{r}^n) \cdot (\delta u (\mathbf{r}^n))^\alpha\rangle_{\mathbf{R}^N}.\tag{S2}
\end{equation}
It follows from Eq. (S2) that the entropic force can be expressed as 
\begin{equation}
    \mathcal{S}_I = -\frac{1}{k_BT^2} \cdot M_I(1)\tag{S3}
\end{equation}
Similarly, it can be shown from Eqs. (S1) and (S2) that the heat capacity force is then
\begin{equation}
    \mathcal{C}_I = \frac{2}{k_BT^2} \cdot M_I(1) - \frac{1}{k_B^2T^3} \cdot M_I(2).\tag{S4}
\end{equation}
We next provide a proof by induction that all terms in the Taylor series of the PMF force, that is, all $\dfrac{\partial^n \mathcal{F}}{\partial T^n}$ can be expressed in terms of the moments provided by Eq. (S2).  
We prove by induction that 
\begin{equation}
    \dfrac{\partial^n \mathcal{F}_I}{\partial T^n} = \sum_{\alpha = 1}^n c_{n,\alpha}\cdot M_I(\alpha).\tag{S5}
\end{equation}
Eq. (S3) constitutes the base case for the induction proof. We next assert that Eq. (S5) holds. Then, for the $n+1$ case, the following holds
\begin{equation}
    \dfrac{\partial}{\partial T}\left (\dfrac{\partial^n \mathcal{F}_I}{\partial T^n} \right )= \sum_{\alpha = 1}^n \dfrac{dc_{n,\alpha}}{dT}\cdot M_I(\alpha) + \sum_{\alpha = 1}^n c_{n,\alpha}(T)\cdot \dfrac{\partial M_I(\alpha)}{\partial T},\tag{S6}
\end{equation}
however, it can be shown that
\begin{equation}
    \dfrac{\partial M_I(\alpha)}{\partial T} = -\frac{\alpha}{T}M_I(\alpha) + \frac{1}{k_BT^2} M_I(\alpha+1),\tag{S7}
\end{equation}
which concludes the proof by induction. Eq. (S5) implies that higher order terms in the Taylor series comprise increasingly larger energy-force moments. Practically, the estimations of these moments are likely to possess a larger variance as one progresses further into the series. Consequently, we do not explore in this work contributions that are larger than second order. Importantly, all energetic components of this series are weighted by the energy factor $1/k_BT$. Therefore, the convergence of the Taylor series is determined by whether the energy-force moments diminish in magnitude relative to $k_BT$.
\section{Coarse-Grained Modeling and Simulation}
The CG potentials developed consisted of a HIP-NN-TS model and a repulsive prior term to discourage sampling of unphysical regions, i.e., near zero regions of the RDF. 
A repulsive prior term of the form $Ae^{-\alpha r}$ was employed, parameter values were assigned according to [9]. 
The hyperparameters and evaluation metrics for all HIP-NN-TS models are shown in Tables ~\ref{tab:hyperparameters} and ~\ref{tab:metric}, respectively. 
Training was conducted on 100 CG frames for all models with a training:validation:test ratio of 0.8:0.1:0.1. 
Loss functions consisted of equally-weighted root mean squared error (RMSE) and mean average error (MAE) for PMF, entropic, and heat capacity forces. 
A batch size of 1 and a learning rate of $1 \times 10^3$ were chosen, and training was conducted for up to a tolerance of 20 epochs without improved validation loss. 
Heat capacity energies and forces were weighted equally during training.

All MD simulations were conducted with the LAMMPS software package.[29] 
Trajectory analysis was performed with the OpenMSCG and mdanalysis software packages.[30,31] 
Mean force calculations on CG frames were obtained from 0.875 ns of restrained MD simulations. 
A restraint constant of 200 kcal/mol $\text{\r{A}}^2$  and a timestep of 0.2 fs were employed. 
Temperature was maintained with a Nose-Hoover thermostat with a damping parameter of 100 timesteps. 
CG simulations were ran for 200000 CG timesteps, with a timestep of 1 fs and Nose-Hoover damping parameter of 1000 timesteps. 

\begin{figure*}[t]
  \includegraphics[width=\columnwidth]{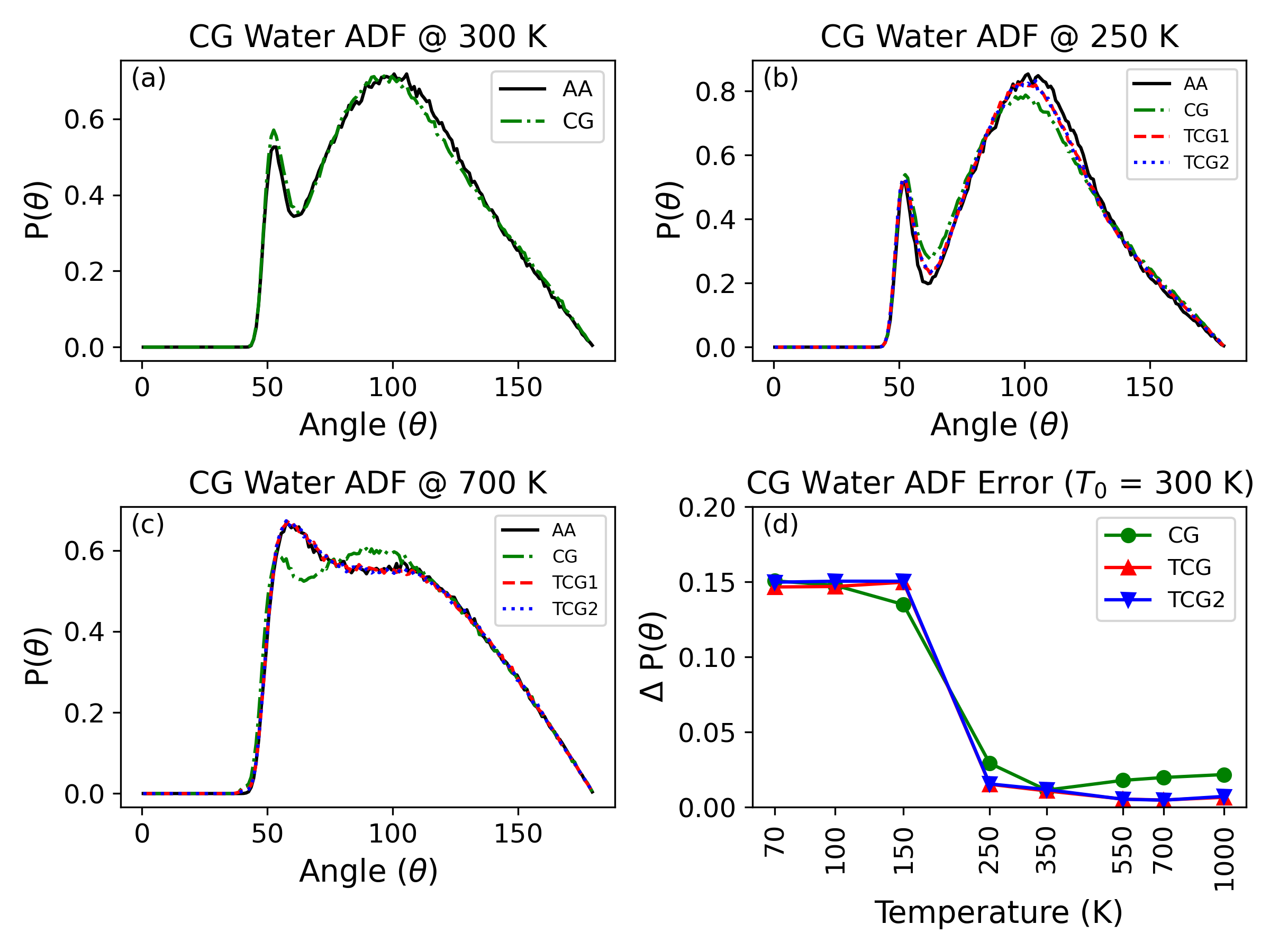}
  \caption{ADF behavior of developed CG models ($T_0 = 300$ K). ADFs of mapped AA and CG models at (a) trained temperature, (b) T = 250 K, and (c) T = 700 K are shown. (d) ADF error across simulated temperatures for all CG models.}
  \label{fig:adf}
\end{figure*}

\begin{figure*}[t]
  \includegraphics[width=\columnwidth]{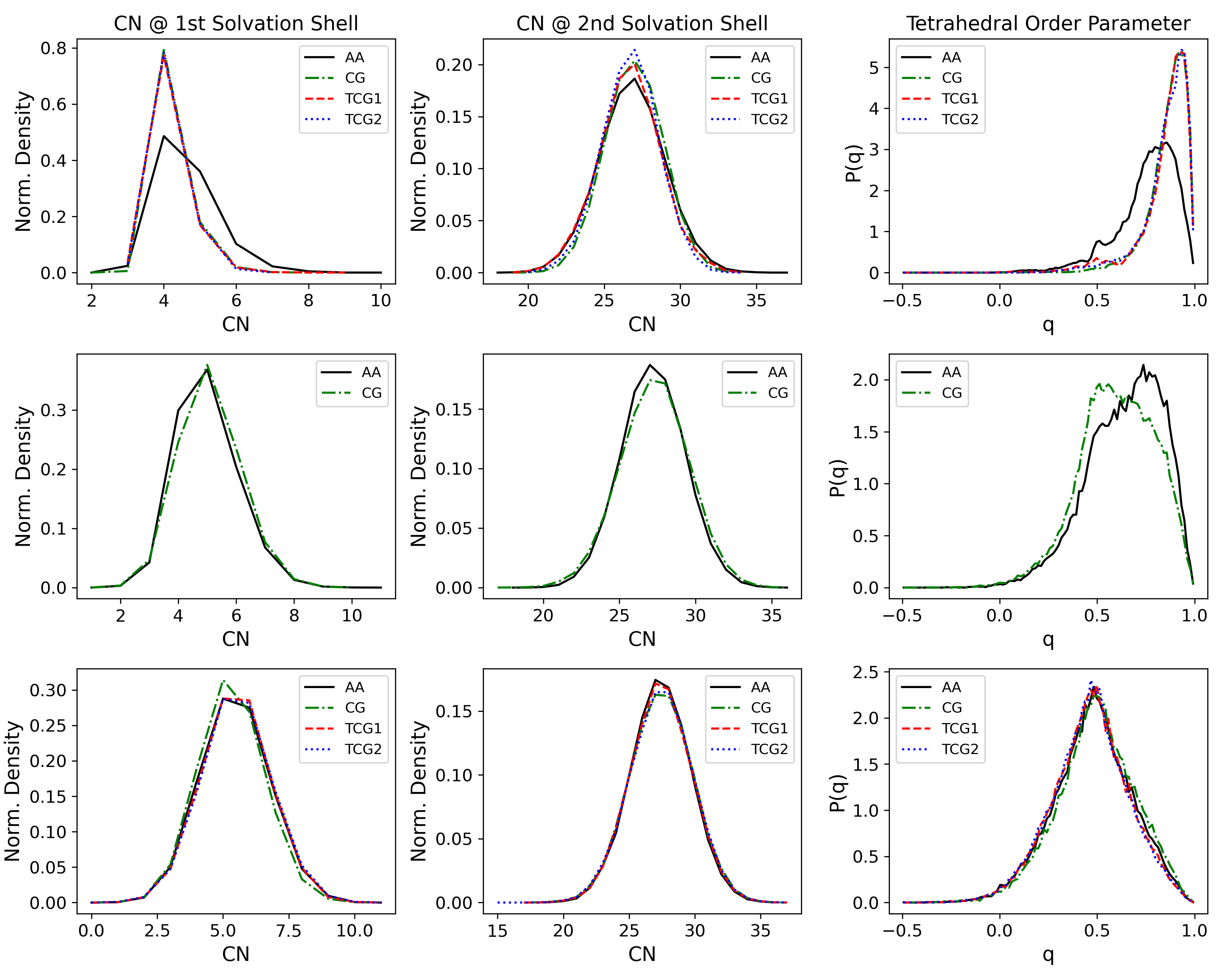}
  \caption{Local environments for mapped AA and CG models developed at $T_0$ = 300 K. Temperatures include 70 K (upper row), 300 K (middle row), and 550 K (lower row).}
  \label{fig:ops}
\end{figure*}

\begin{figure*}[t]
  \includegraphics[width=\columnwidth]{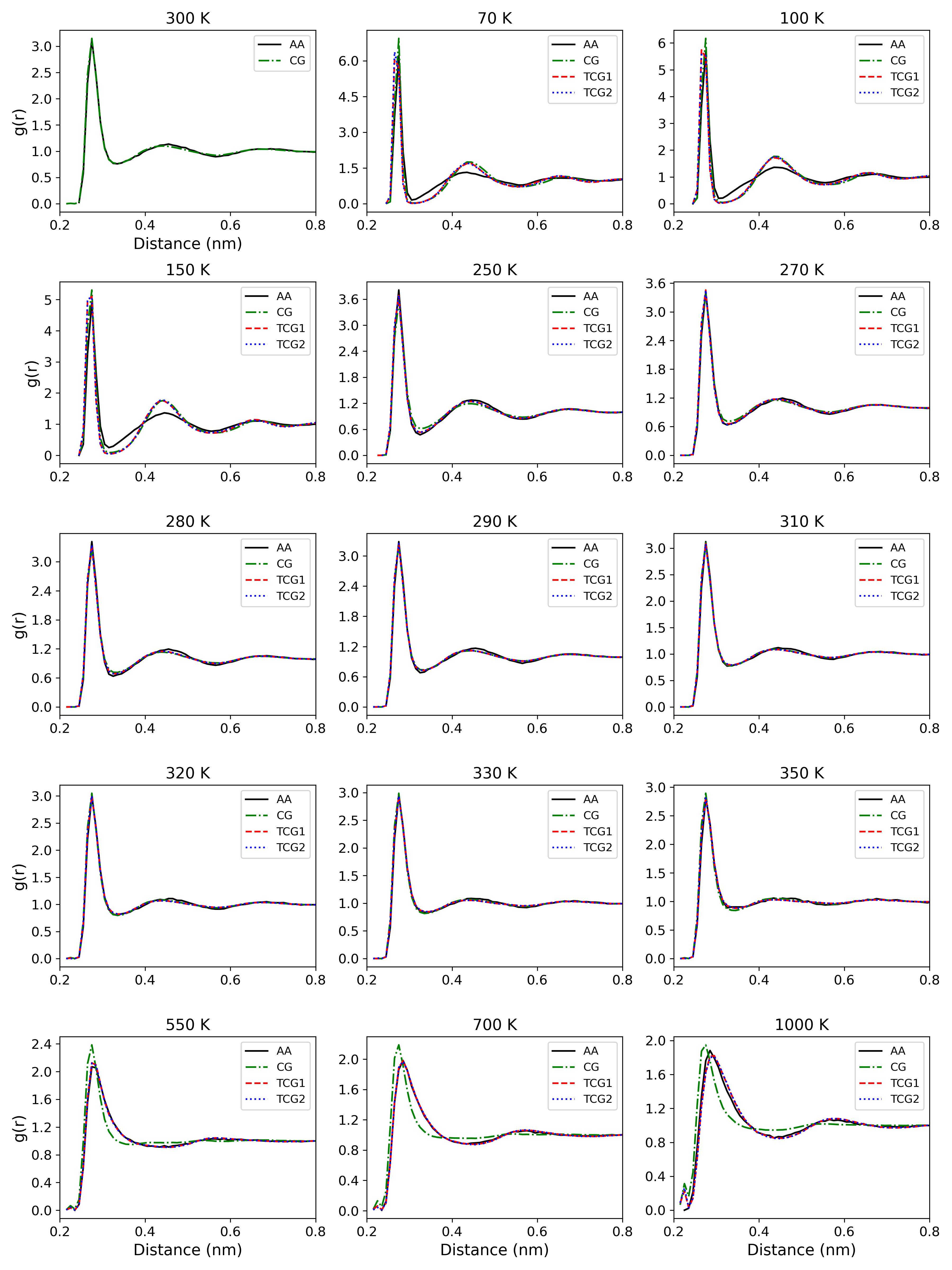}
  \caption{RDF plots of AA and CG models for $T_0$ = 300 K and at temperatures not observed during training.}
  \label{fig:rdf300}
\end{figure*}

\begin{figure*}[t]
  \includegraphics[width=\columnwidth]{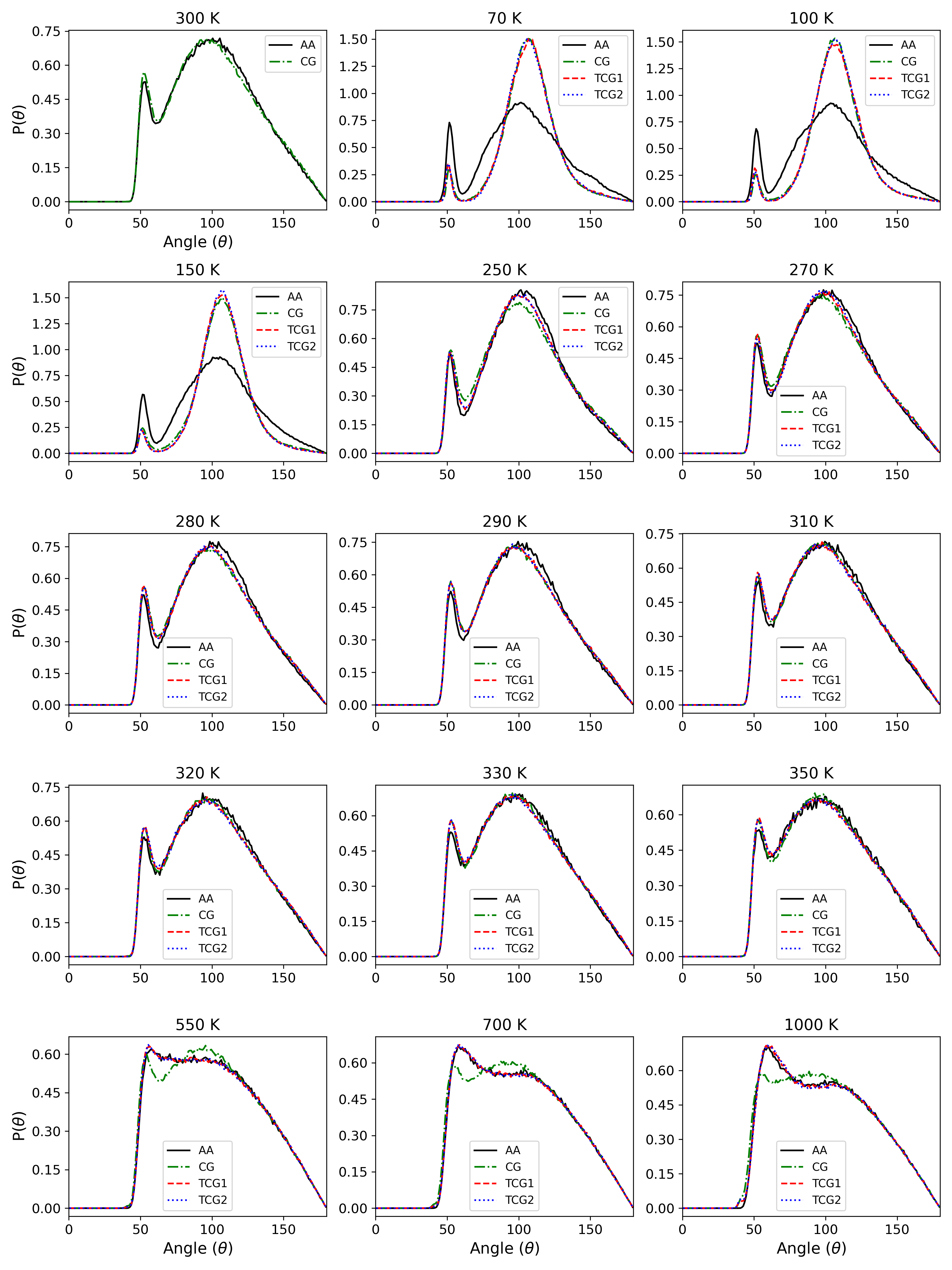}
  \caption{ADF plots of AA and CG models for $T_0$ = 300 K and at temperatures not observed during training.}
  \label{fig:adf300}
\end{figure*}

\begin{figure*}[t]
  \includegraphics[width=\columnwidth]{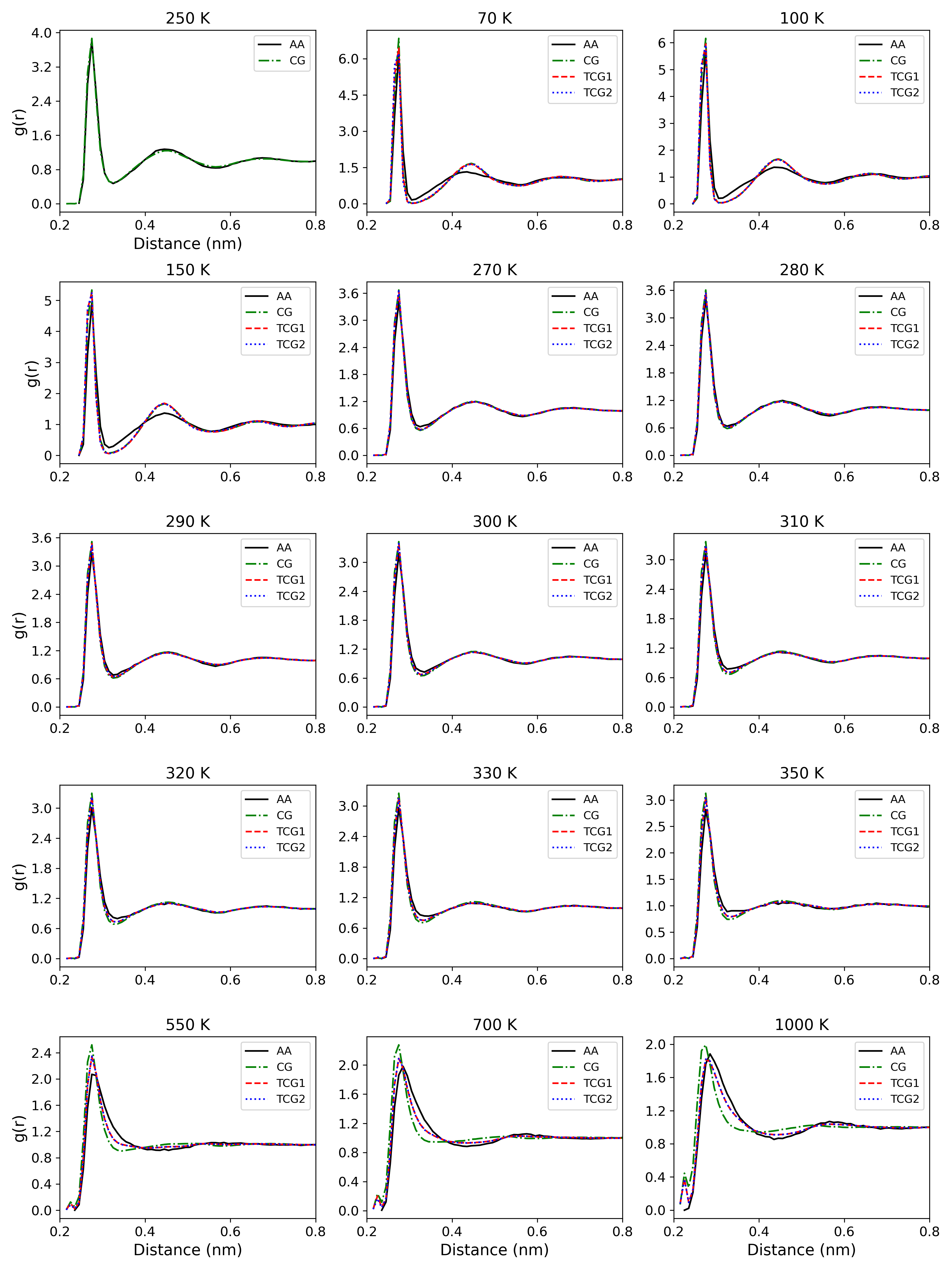}
  \caption{RDF plots of AA and CG models at $T_0$ = 250 K and at temperatures not observed during training.}
  \label{fig:rdf250}
\end{figure*}

\begin{figure*}[t]
  \includegraphics[width=\columnwidth]{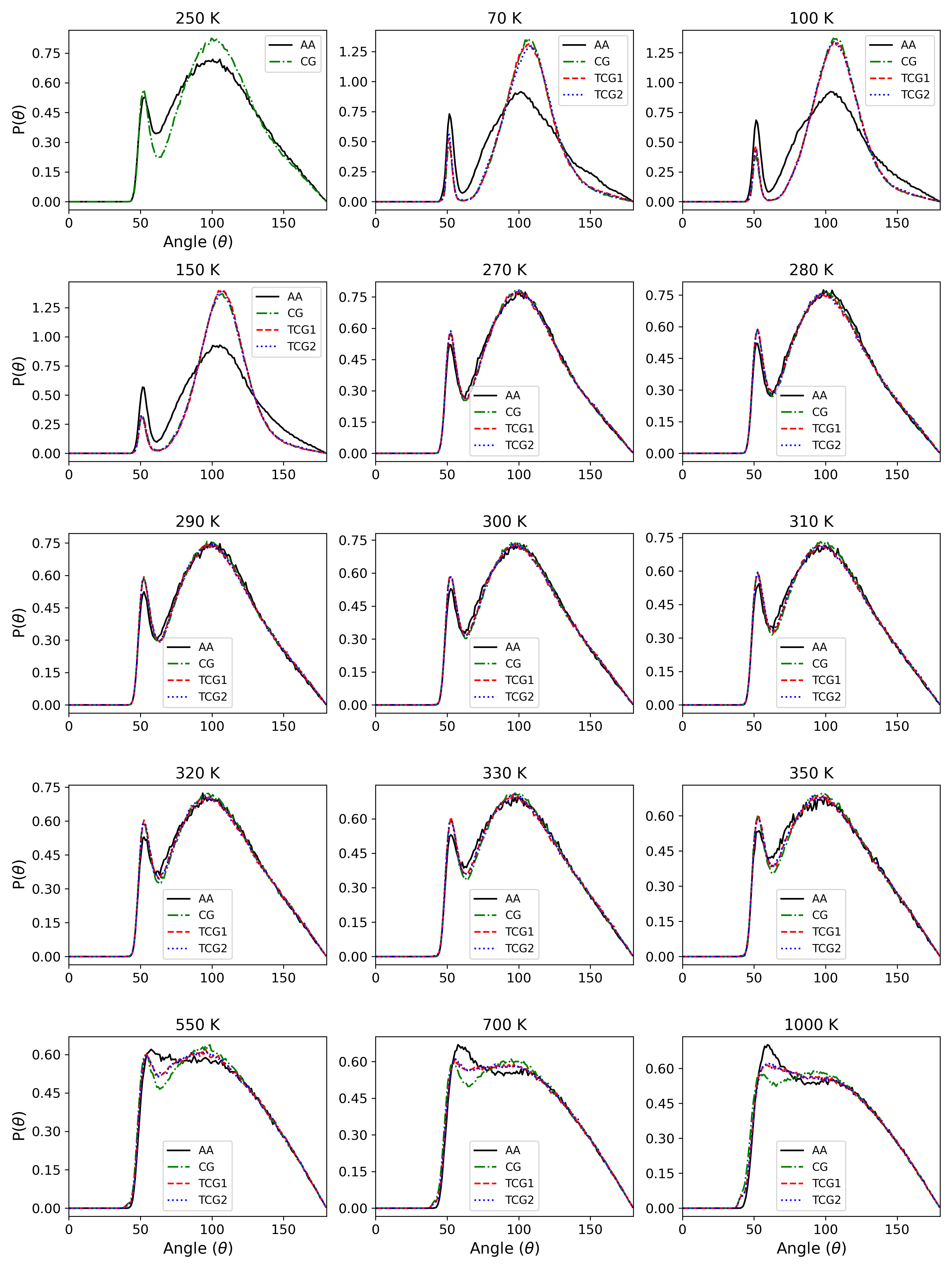}
  \caption{ADF plots of AA and CG models at $T_0$ = 250 K and at temperatures not observed during training.}
  \label{fig:adf250}
\end{figure*}

\begin{figure*}[t]
  \includegraphics[width=\columnwidth]{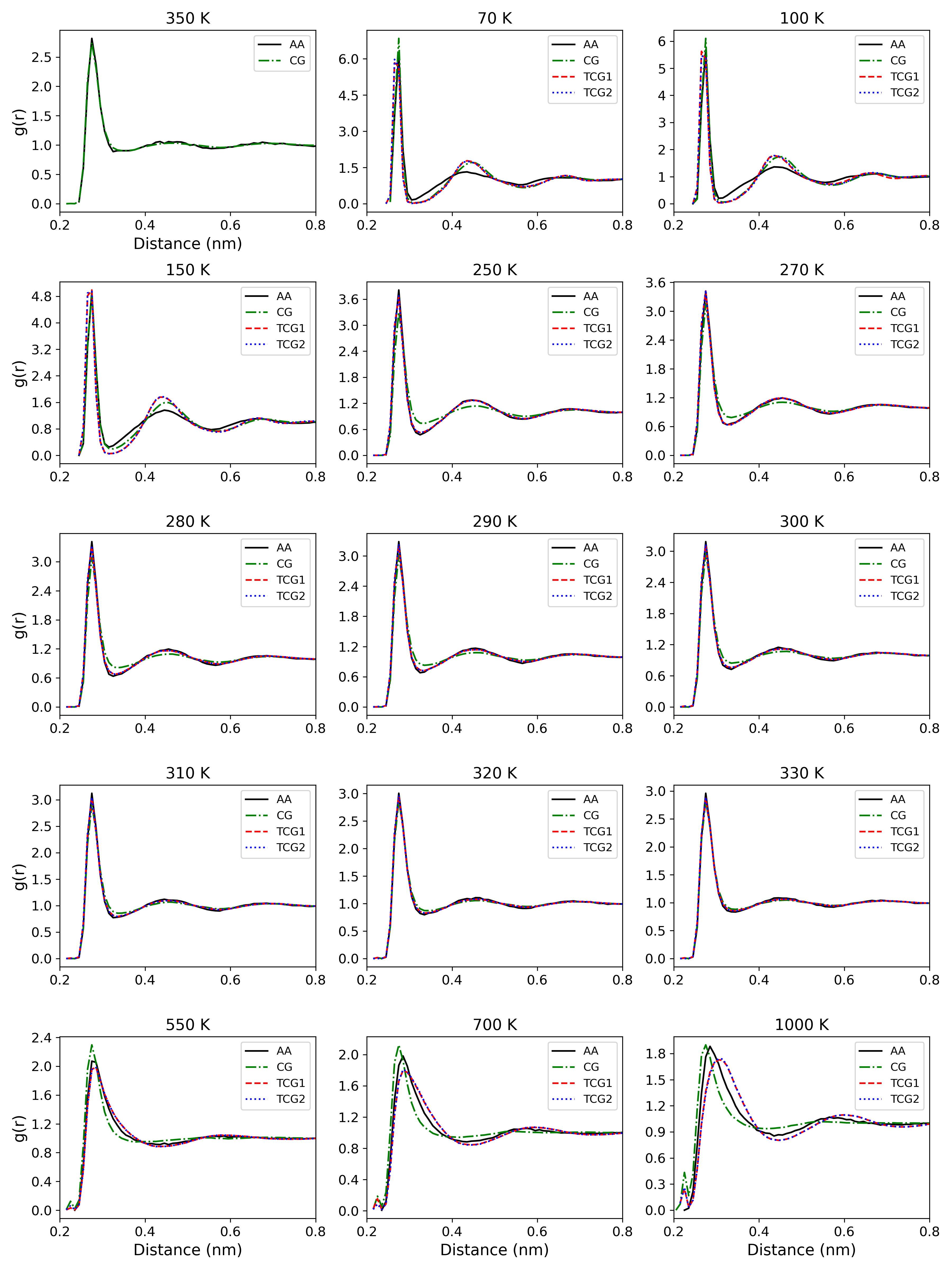}
  \caption{RDF plots of AA and CG models at $T_0$ = 350 K and at temperatures not observed during training.}
  \label{fig:rdf350}
\end{figure*}

\begin{figure*}[t]
  \includegraphics[width=\columnwidth]{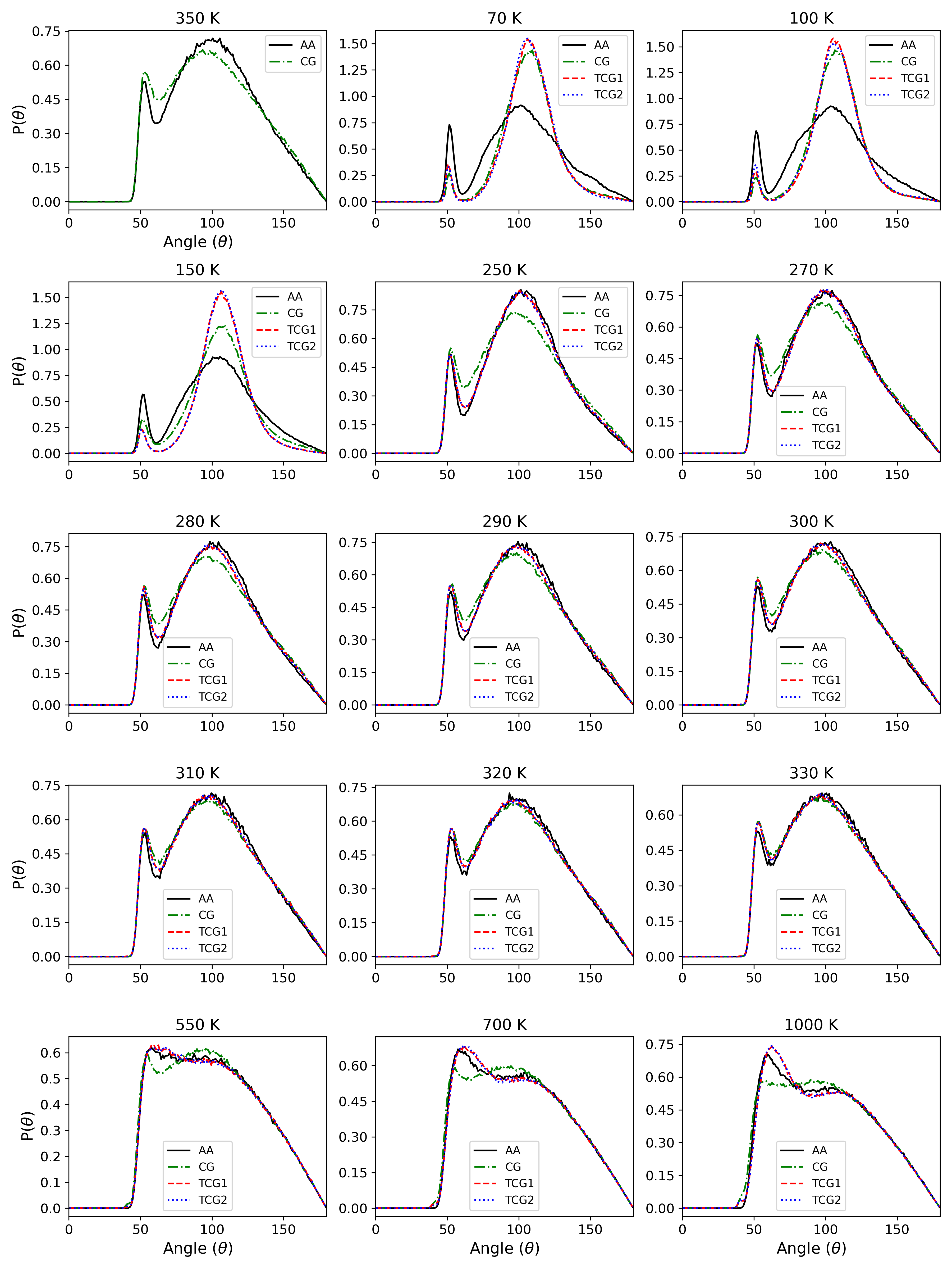}
  \caption{ADF plots of AA and CG models at $T_0$ = 350 K and at temperatures not observed during training.}
  \label{fig:adf350}
\end{figure*}

Representative ADF plots and errors are shown for all CG models developed at $T_0$ = 300 K in Fig.~\ref{fig:adf}. 
Local environments for representative temperatures are shown in Fig.~\ref{fig:ops}.
RDF plots and ADF plots for $T_0 = 300 K$ for all temperatures investigated in this work are shown in Figs.~\ref{fig:rdf300} and Figs.~\ref{fig:adf300}.
RDF plots and ADF plots for $T_0 = 250 K$ and $T_0 = 300 K$ are also shown in Fig.~\ref{fig:rdf250}, Fig.~\ref{fig:adf250}, Fig.~\ref{fig:rdf350}, andd Fig.~\ref{fig:adf350}, respectively.

\begin{table}
  \caption{Hyperparameters for ML CG FFs.}
  \label{tab:hyperparameters}
    \begin{tabular}{ll}
        Hyperparameters  & Value \\
        $\verb|n_features|$  & 128\\
        $\verb|n_sensitivities|$ & 20 \\
        $\verb|dist_soft_min|$ & 2.2 \r{A}\\
        $\verb|dist_soft_max|$ & 9.0 \r{A}\\
        $\verb|dist_hard_max|$ & 10.0 \r{A}\\ 
        $\verb|n_interaction_layers|$ & 1\\
        $\verb|n_atom_layers|$ & 3\\
        $\verb|sensitivity_type|$ & inverse \\
        $\verb|resnet|$ & true
        \end{tabular}
\end{table}

\begin{table}
  \caption{Loss metrics for ML CG FFs trained.}
  \label{tab:metric}
  \begin{ruledtabular}
    \begin{tabular}{cccc}
      Metric & Test Loss ($T_0$ = 250 K) & Test Loss ($T_0$ = 300 K) & Test Loss ($T_0$ = 350 K)  \\
      \hline
        RMSE($\mathcal{F}$) (kcal/mol \r{A}) &  2.154 $\pm$ 0.057 & 2.317 $\pm$ 0.037 & 2.118 $\pm$ 0.028 \\
        MAE($\mathcal{F}$) (kcal/mol \r{A}) & 1.659 $\pm$ 0.042 & 1.795 $\pm$ 0.029 & 1.654 $\pm$ 0.024 \\
        $R^2$($\mathcal{F}$) & 0.821 $\pm$ 0.009 & 0.809 $\pm$ 0.006 & 0.835 $\pm$ 0.004 \\
        RMSE($\mathcal{S}$) (kcal/mol \r{A}) & 1.352 $\pm$ 0.002 &  4.042 $\pm$ 0.003 & 11.129 $\pm$ 0.002 \\
        MAE($\mathcal{S}$) (kcal/mol \r{A}) & 1.015 $\pm$ 0.002 & 2.246 $\pm$ 0.004 & 6.720 $\pm$ 0.005 \\
        $R^2$($\mathcal{S}$) &  0.146 $\pm$ 0.003 & 0.087 $\pm$ 0.001 &  0.029 $\pm$ 0 \\
        RMSE($C_V$) (kcal/mol) & 90.570 $\pm$ 0.503  & 179.917 $\pm$ 3.528 & 193.693 $\pm$ 2.492  \\
        MAE($C_V$) (kcal/mol) &  70.156 $\pm$ 0.317 & 153.767 $\pm$ 2.755 & 178.393 $\pm$ 3.159 \\
        $R^2$($C_V$) & -0.041 $\pm$ 0.012  & -0.212 $\pm$ 0.047 &  -0.164 $\pm$ 0.030 \\
        RMSE($\mathcal{C}$) (kcal/mol \r{A}) &  297.933 $\pm$ 0.003 & 809.650 $\pm$ 0 & 1745.567 $\pm$ 0.033 \\
        MAE($\mathcal{C}$) (kcal/mol \r{A}) & 230.943 $\pm$ 0.003  & 511.280 $\pm$ 0 & 1107.400 $\pm$ 0 \\
        $R^2$($\mathcal{C}$) & 4.21e-05 $\pm$ 2.58e-05  & 6.43e-05 $\pm$ 3.91e-07 &  6.58e-05 $\pm$ 2.02e-06
    \end{tabular}
  \end{ruledtabular}
\end{table}

\section{Energy-Entropy Decomposition of the Coarse-Grained Potential of Mean Force}
\begin{figure*}[t]
  \includegraphics[width=\columnwidth]{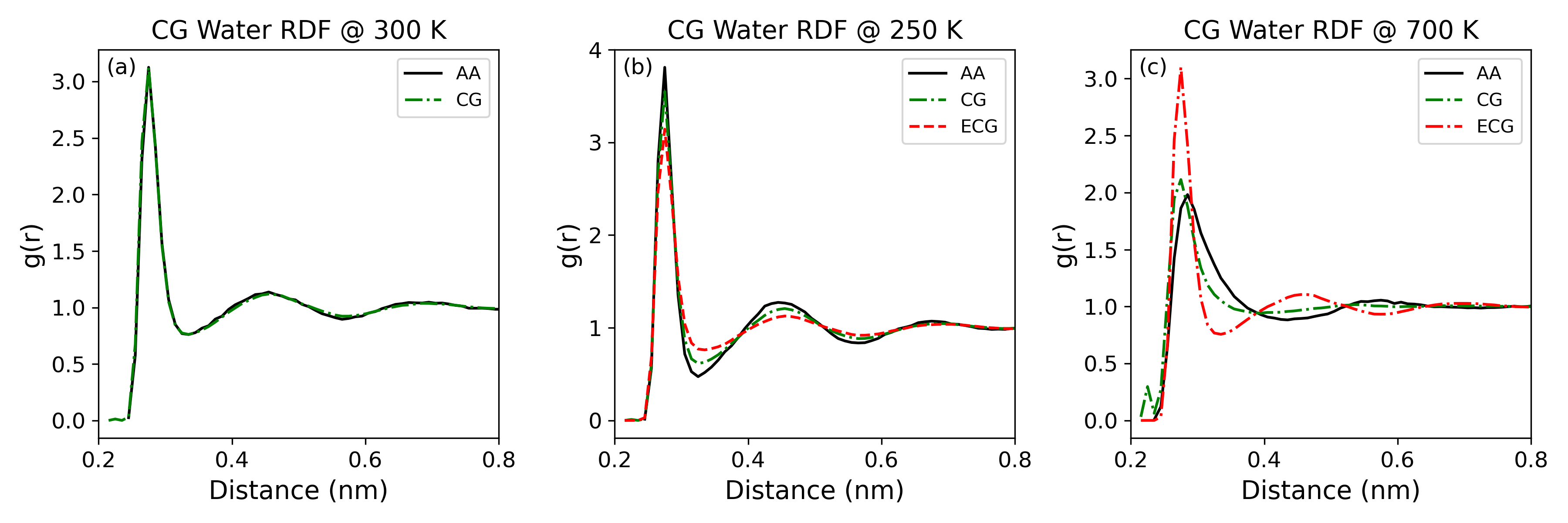}
  \caption{Behavior of CG models developed via energy-entropy decomposition at $T_0$ = 300 K. RDFs at (a) reference temperature, (b) 250 K, and (c) 700 K are shown.}
  \label{fig:ecg}
\end{figure*}

We discuss and compare a previously established alternative to incorporating temperature dependence into CG models as a $1^\text{st}$ order correction. We define the CG internal energy, $U(\mathbf{R}^N;T)$ as 
\begin{equation}\label{Eq. S8}
    U(\mathbf{R}^N;T) = \langle u(\mathbf{r}^n)\rangle_{\mathbf{R}^N, T}.\tag{S8}
\end{equation}
The CG internal energy can in principle be learned from regression. For a CG model $\tilde{U}(\mathbf{R}^N)$ with tunable parameters $\boldsymbol{\omega}$, we define the following loss
\begin{equation}\label{Eq. S9}
    \mathcal{L}_{ME}(\boldsymbol{\omega}) = \Big \langle \| U(\mathbf{R}^N) - \tilde{U}(\mathbf{R}^N; \boldsymbol{\omega})\|^2 \Big \rangle.\tag{S9}
\end{equation}
To first order, the free energy at a given temperature can be approximated as
\begin{equation}\label{Eq. S10}
    F(\mathbf{R}^N;T) = \frac{T}{T_0} \cdot F(\mathbf{R}^N;T_0) + \left ( 1 - \frac{T}{T_0}\right ) \cdot U(\mathbf{R}^N; T_0) \tag{S10}.
\end{equation}
Eqs. (S9) and (S10) provide an energetic approach to learning $1^\text{st}$ order transferability, in contrast to the entropic approach depicted in this work. 
We refer to models developed in this manner as ECG models. 
A key difference in the practical implementation of the two approaches is that learning from thermal response forces is inherently local, providing $3N$ regression targets for a given constrained CG configuration, while learning from internal energy is by default global, i.e., only $1$ regression target is provided from a given CG configuration. 
Consequently, for finite datasets the two approaches are expected to differ, even though they are in theory identical at first order. Fig. ~\ref{fig:ecg} demonstrates the clear difference in the resulting CG model behavior for CG models developed from energy-matching and thermal response force-matching. 
We attribute the significantly improved accuracy of CG models developed by thermal response force-matching to the local rather than global regression targets employed for training.

\section{Predictive Coarse-Grained Dynamics from Temperature Transferability}

\begin{figure*}[t]
  \includegraphics[width=\columnwidth]{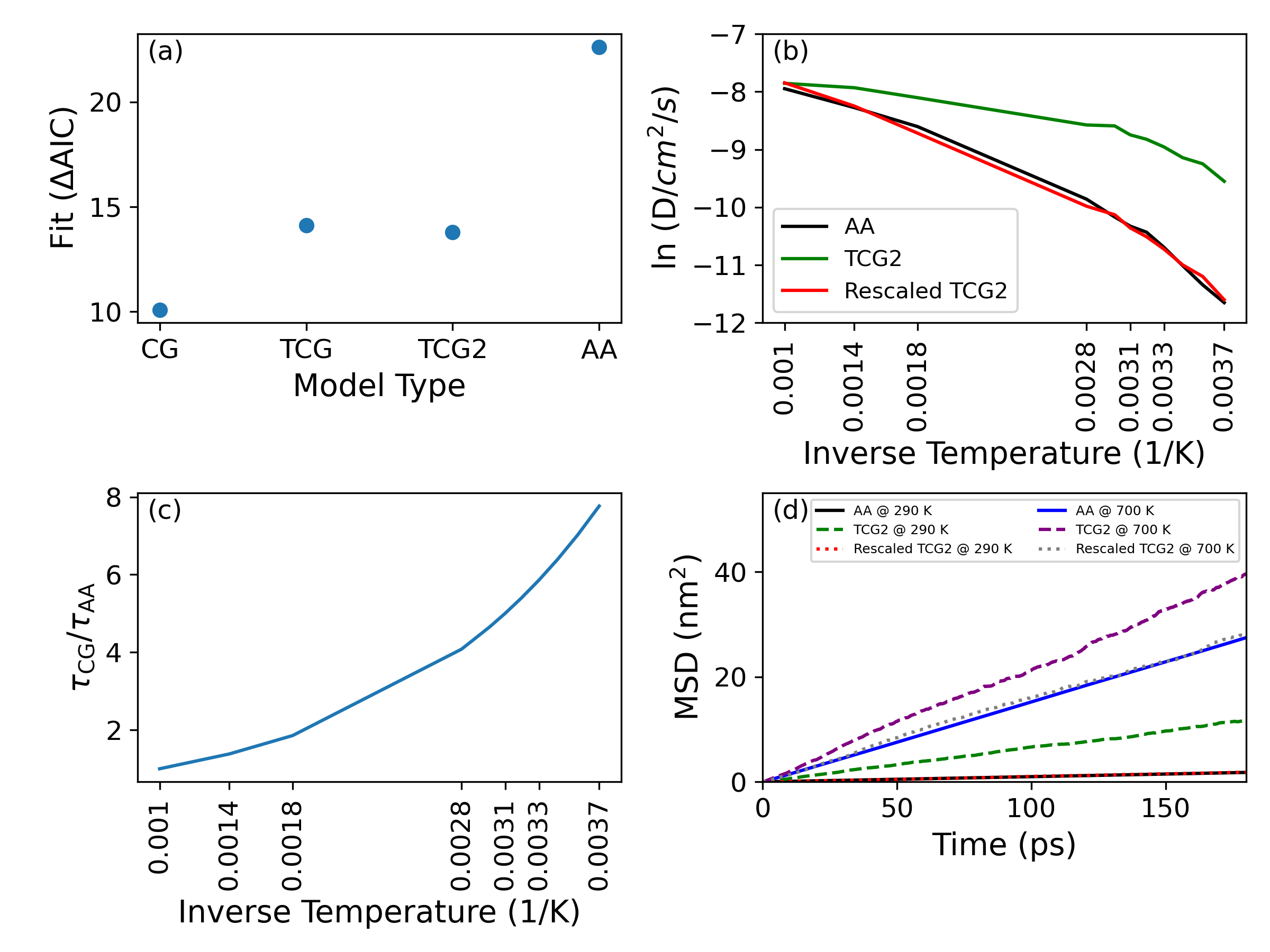}
  \caption{(a) $\Delta$AIC is shown for all models. (b) Diffusion plots are shown for the mapped AA model, the TCG2 model, as well as the rescaled diffusion of the TCG2 model. (c) A plot of the amount of AA time covered within a CG timestep is shown. (d) Mean-squared distance plots of AA and TCG2 model are shown. The time of the TCG2 simulation is also rescaled and its resulting MSD is shown.}
  \label{fig:dyn}
\end{figure*}

We quantify the degree of non-Arrhenius behavior for all models according to the Akaike Information Criterion[32] (AIC) between a linear ($\text{AIC}_1$) and quadratic ($\text{AIC}_2$) fit of $\log D(T)$ vs $T$, $\Delta $\text{AIC} = $\text{AIC}_1 - \text{AIC}_2$.
In general, we find that addition of temperature correction terms produces a larger $\Delta \text{AIC}$ relative to the temperature-independent CG model, with the mapped AA model showing the largest deviation from Arrhenius behavior, as shown in Fig.~\ref{fig:dyn}(a).
We next elaborate on how the CG models developed possess predictive capability over CG dynamics. Our primary assumption is that diffusion at the CG resolution is well-described via overdamped Kramers' theory,
\begin{equation}\label{Eq. S11}
    D(T)= \frac{\omega_{\text{max}}(T)}{\gamma(T)}\frac{\omega_{\text{min}}(T)}{2\pi} e^{-\beta F(T)^\dagger/k_BT},\tag{S11}
\end{equation}
where $\omega_{\text{max}}$ is the curvature at the top of the transition and $\omega_{\text{min}}$ is the curvature at the minimum of the transition.
Crucially, both of these terms as well as the exponential factor are reliant only on the CG PMF. The dynamical disparity is then attributed to the friction, $\gamma(T)$.
Following the main text, we assume a fit of the ratio of diffusion constants between mapped AA and CG models as
\begin{equation}\label{Eq. S12}
    \ln \left (\frac{D_{\text{AA}}}{D_{\text{CG}}} \right )= -\alpha \cdot T + \beta.\tag{S12}
\end{equation}
Eq. (S12) suggests that the exact diffusion can be predicted from the diffusion of the CG PMF.
Additionally, this suggests, via Eq. (S11), that the friction constants of the mapped AA and CG PMF are related by an exponential factor. 
We demonstrate the predictive capabilities of this fit in Fig.~\ref{fig:dyn}(b).
Furthermore, following the proportionality between diffusion and mean-squared distance, a relationship between the CG and AA timescales can be expressed as
\begin{equation}\label{Eq. S13}
    \tau_{\text{AA}} = \tau_{\text{CG}} \cdot e^{-\alpha \cdot T + \beta}.\tag{S13}
\end{equation}
In Fig.~\ref{fig:dyn}(c) and Fig.~\ref{fig:dyn}(d), we show how the ratio of these timescales varies with $T$ according to Eq. (S10) and demonstrate how this time rescaling can be employed to deduce accurate CG dynamics from CG simulations, respectively. It is clear that as $T$ decreases, CG time accounts for an increasingly larger period of AA time. This can at least partially be attributed to the prevalence of hydrogen bonding networks in the lower $T$ regime, which are fully described by the AA model but which must emerge in the CG model as friction.


\end{document}